\documentclass[12pt]{article}
\usepackage{amsmath}
\usepackage{graphicx,psfrag,epsf}
\usepackage{enumerate}
\usepackage{natbib}
\usepackage{url} % not crucial - just used below for the URL 
\usepackage{multirow}
\usepackage{color}

\newcommand{\blind}{0}

\begin{document}

\def\spacingset#1{\renewcommand{\baselinestretch}%
{#1}\small\normalsize} \spacingset{1}

\newcommand{\pkg}[1]{{\fontseries{b}\selectfont #1}} 

%%%%%%%%%%%%%%%%%%%%%%%%%%%%%%%%%%%%%%%%%%%%%%%%%%%%%%%%%%%%%%%%%%%%%%%%%%%%%%

\if0\blind
{
  \title{\bf FFT-Based Fast Computation of Multivariate Kernel Density Estimators with Unconstrained Bandwidth Matrices}
  \author{Artur Gramacki \\
  %\thanks{
  %  The authors gratefully acknowledge \textit{please remember to list all relevant funding sources in the unblinded version}}\hspace{.2cm}\\
    Institute of Control and Computation Engineering\\
    University of Zielona G\'o{}ra\\
    ul. Licealna 9, Zielona G\'o{}ra 65-417, Poland\\
    E-mail: a.gramacki@issi.uz.zgora.pl\\
    and \\
    Jaros\l{}aw Gramacki \\
    Computer Center\\
    University of Zielona G\'o{}ra\\
    ul. Licealna 9, Zielona G\'o{}ra 65-417, Poland\\
    E-mail: j.gramacki@ck.uz.zgora.pl}
  \maketitle
} \fi

\if1\blind
{
  \bigskip
  \bigskip
  \bigskip
  \begin{center}
    {\LARGE\bf Title}
\end{center}
  \medskip
} \fi

\bigskip
\begin{abstract}
The problem of fast computation of multivariate kernel density estimation (KDE) is still an open research problem. In our view, the existing solutions do not resolve this matter in a satisfactory way. One of the most elegant and efficient approach utilizes the fast Fourier transform. Unfortunately, the existing FFT-based solution suffers from a serious limitation, as it can accurately operate only with the constrained (i.e., diagonal) multivariate bandwidth matrices. In this paper we describe the problem and give a satisfactory solution. The proposed solution may be successfully used also in other research problems, for example for the fast computation of the optimal bandwidth for KDE.
\end{abstract}

\noindent%

\vfill

\newpage
\spacingset{1.45} % DON'T change the spacing!
\section{Introduction}\label{sec-intro}
Kernel density estimation is one of the most important statistical tool with many practical applications, see for example \cite{Kulczycki:2010,Duong:2013} and many others. It has been applied successfully to both univariate and multivariate problems. There exists extensive literature on this issue, including several classical monographs, see \cite{Silverman:1998}, \cite{Scott:1992}, \cite{Simonoff:1996} and \cite{Wand-1995}. A general form of the $d$-dimentional multivariate kernel density estimator is

\begin{align}\label{eq-kde}
\hat{f}(\boldsymbol{x},\boldsymbol{H}) 
&= 
\frac{1}{n} 
\sum_{i=1}^n 
K_{\boldsymbol{H}} \left( \boldsymbol{x}-\boldsymbol{X}_i \right),  \nonumber \\
K_{\boldsymbol{H}}(u) 
&= 
|\boldsymbol{H}|^{-1/2}K \left(\boldsymbol{H}^{-1/2}u\right)
\end{align}
where
$\boldsymbol{H}$ is the symmetric and positive definite $d \times d$ matrix (called \emph{bandwidth} or \emph{smoothing} matrix), $d$ is the problem dimensionality, $\boldsymbol{x}=(x_1, x_2, \cdots, x_d)^T$, 
$\boldsymbol{X}_i=(X_{i1}, X_{i2}, \cdots, X_{id})^T$, $i=1,2,\cdots,n$ is a sequence of independent identically distributed (iid) $d$-variate sample drawn from some distribution with an unknown density $f$,
$K$ and $K_{\boldsymbol{H}}$ are the unscaled and scaled kernels, respectively. In most cases the kernel has the form of a standard multivariate normal density.

There are two main computational problems related to KDE: (a) fast evaluation of the kernel density estimate $\hat{f}(\boldsymbol{x},\boldsymbol{H})$, (b) fast estimation of the optimal bandwidth matrix $\boldsymbol{H}$ (or scalar $h$ in the univariate case). In the paper we concentrate on the first problem. 

It is obvious from Eqn.~(\ref{eq-kde}) that the naive direct evaluation of the KDE at $m$ evaluation points for $n$ data points requires $O(mn)$ kernel evaluations. Evaluation points can be of course the same as data points and then the computational complexity is $O(n^2)$ making it very expensive, especially for large datasets and higher dimensions. 

A number of methods have been proposed to accelerate the computations. See for example \cite{Raykar:2010} for an interesting review of the methods. Other techniques, like for example usage of Graphics Processing Units (GPUs) are also used \citep{Andrzejewski:2013}. One of the most elegant and effective methods is based on using the fast Fourier transform (FFT). A preliminary work on using FFT to kernel density estimation was given in \cite {Silverman:1982}(only for univariate case). 

In the paper we are concerned with an FFT-based method that was originally described by \cite{Wand-1994}. In \citet[appendix D]{Wand-1995} an illustrative toy example has been presented. The method works very well for univariate case but, unfortunately, its multivariate extension does not support unconstrained bandwidth matrices.
From now on this method will be called \emph{Wand's algorithm}. The FFT-based method investigated in this paper can be easily adapted also in other algorithms, for example for the fast computation of the optimal bandwidth for KDE. An appropriate research paper is in preparation and its draft version can be found in \cite{Gramacki:2015}.

The remainder of the paper is organized as follows: in Section \ref{sec-algo} we briefly present the FFT-based algorithm and indicate its limitations. in Section \ref{sec-demo} we demonstrate the problem. In Section \ref{sec-ident} we identify the source of the problem and propose a satisfactory solution. In Section \ref{sec-conlusion} we conclude the paper.

\section{FFT-based algorithm for KDE}\label{sec-algo}

Below we briefly present Wand's algorithm. It consists of 3 basic steps. In the \textbf{first step} the multivariate  \emph{linear binning} (a kind of data discretization, see \cite{Wand-1994}) of the input random variables $\boldsymbol{X}_i$ is required. The binning approximation of Eqn. (\ref{eq-kde}) is 
\begin{align}\label{eq-bkde}
\tilde{f}(\boldsymbol{g_j}, \boldsymbol{H}, \boldsymbol{M}) = 
\frac{1}{n}
\sum_{l_1=1}^{M_1} 
\cdots
\sum_{l_d=1}^{M_d} 
K_{\boldsymbol{H}} \left(\boldsymbol{g_j} - \boldsymbol{g_l} \right) \boldsymbol{c_l}
\end{align}
where $\boldsymbol{g}$ are equally spaced \emph{grid points} and $\boldsymbol{c}$ are \emph{grid counts}. Grid counts are obtained by assigning certain weights to the grid points, based on neighbouring observations. In other words, each grid point is accompanied by a corresponding grid count. 

The following notation is used: for $k=1,\ldots,d$, let $g_{k1} < \cdots < g_{kM_K}$ be an equally spaced grid in the $k$th coordinate directions such that $[g_{k1}, g_{kM_k}]$ contains the $k$th coordinate grid points. Here $M_k$ is a positive integer representing the \emph{grid size} in direction $k$. Let
\begin{align}
\boldsymbol{g_j} =
(g_{1j_1}, \ldots, g_{dj_d}), \;\;\; 1 \le j_k \le M_k, \;\;\; k=1,\ldots, d
\end{align}
denote the grid point indexed by $\boldsymbol{j}=(j_1,\ldots,j_d)$ and the $k$th binwidth  be denoted by
\begin{align}\label{eq-delta-k}
\delta_k = \frac{g_{kM_k} - g_{k1}} {M_k - 1}.
\end{align}

In the \textbf{second step} Eqn. (\ref{eq-bkde}) is rewritten so that it takes a form of the \emph{convolution}
\begin{align}\label{eq-bkde-conv}
\tilde{f}_{\boldsymbol{j}} 
= 
\sum_{l_1=-(M_1-1)}^{M_1-1} 
\cdots 
\sum_{l_d=-(M_d-1)}^{M_d-1} 
\boldsymbol{c_{j-l}} \boldsymbol{k_l}
\end{align}  
where
\begin{align}\label{eq-kl}
\boldsymbol{k_l} 
= 
\frac{1}{n} K_{\boldsymbol{H}} (\delta_1 l_1, \cdots, \delta_d l_d). 
\end{align} 

In the \textbf{third step} we compute the convolution between $\boldsymbol{c_{j-l}}$ and $\boldsymbol{k_l}$ using the FFT algorithm in only $O(M_1 \log M_1 \ldots M_d \log M_d)$ operations compared to the  $O(M_1^2 \ldots M_d^2 )$ operations required for direct computation of Eqn. (\ref{eq-bkde}).

In practical implementations of  Wand's algorithm, the sum limits in $\{M_1, \cdots, M_d\}$ can be additionally  shrunk to some smaller values $\{L_1, \cdots, L_d\}$, which significantly reduces the computational burden. In most cases, the kernel $K$ is the multivariate normal density function and, as such, an \emph{effective support} can be defined, i.e., the region outside which the values of $K$ are practically negligible. Our proposed formula for calculating $L_k, k=1, \cdots, d$ is given in Section~\ref{sec-ident}. Now Eqn. (\ref{eq-bkde-conv}) can be finally rewritten as
\begin{align}\label{eq-bkde-conv-L}
\tilde{f}_{\boldsymbol{j}} 
= 
\sum_{l_1=-L_1}^{L_1} 
\cdots 
\sum_{l_d=-L_d}^{L_d} 
\boldsymbol{c_{j-l}} \boldsymbol{k_l}.
\end{align}

Although the above presented 3-step algorithm is very fast and accurate it suffers from a serious limitation. It supports only a small subset of all possible multivariate kernels of interest. Two commonly used kernel types are \emph{product} and \emph{radial} 
%(also known as \emph{spherically symmetric}) 
ones \citep{Wand-1995}. The problem reveals if the radial kernel is used and the bandwidth matrix $\boldsymbol{H}$ is \emph{unconstrained}, that is $\boldsymbol{H} \in  \mathcal{F}$, where  $\mathcal{F}$ denotes the class of symmetric, positive definite $d \times d$ matrices. If, however, the bandwidth matrix belongs to a more restricted \emph{constrained} (\emph{diagonal}) form (that is $\boldsymbol{H} \in  \mathcal{D}$) the problem doesn't manifest itself. 

To the best of our knowledge, the above mentioned problem is not clearly presented and solved in literature, except a few short mentions in \cite{Wand-1995}, \cite{Wand-1994} and in the \texttt{kde\{ks\}} \textbf{R} function \citep{ks}\footnote{Starting from version 1.10.0 of the \pkg{ks} package, the FFT-based solution presented in this paper was successfully implemented there.}. Moreover, many authors cite the FFT-based algorithm for KDE mechanically, without any qualification or mentioning its greatest limitation. 

\section{Problem demonstration}\label{sec-demo}
\begin{figure}[!hb]
	\centering
	\includegraphics[width=16.5cm]{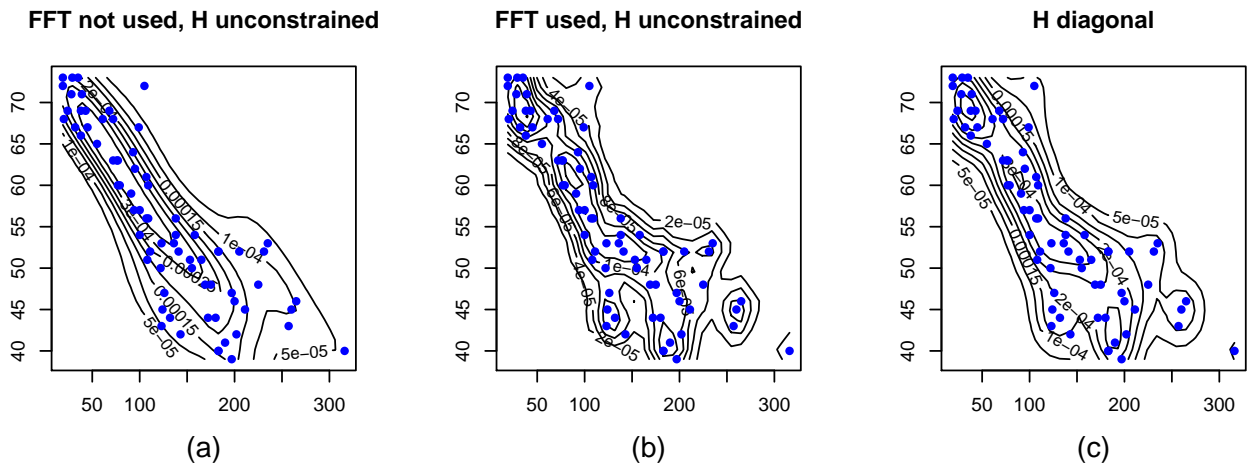}
	\caption{Density estimations for the sample \emph{unicef} dataset with and without using FFT, for both unconstrained and constrained bandwidth matrices. Description of each plot is given in the text.} 
	\label{fig-example}
\end{figure}

In Figure \ref{fig-example} we demonstrate the problem mentioned in Section \ref{sec-algo}. A sample \emph{unicef} dataset from the \pkg{ks} \textsf{R} package was used. For simplicity only 2D examples are shown but extension for higher dimensions is not difficult. For better readability, the authors' own \textsf{R} codes were used and they are provided as supplemental materials\footnote{During experiments a small bug in \texttt{binning\{ks\}} \textsf{R} function was found. According to binning definition, grid counts entries in $\boldsymbol{c_l}$ must sum to $n$. A quick experiment with \texttt{binning\{ks\}} shows that it returns wrong results, while the authors' version returns the correct ones. The corrected version of the binning function is also included in the supplemental materials.}. 
Wand's algorithm is implemented in the \pkg{ks} \textsf{R} package \citep{ks}, as well as  in the \pkg{KernSmooth} \textsf{R} package \citep{KernSmooth}. However, the \pkg{KernSmooth} implementation supports only product kernels. The standard \texttt{density\{stats\}} \textsf{R} function uses FFT to compute univariate kernel density estimates only. 

The density estimation depicted in Figure \ref{fig-example}(a) can be treated as the reference. It was calculated directly according Eqn.~(\ref{eq-bkde}). The bandwidth $\boldsymbol{H}$ was unconstrained and was calculated using the  \texttt{Hpi\{ks\}} \textsf{R} function. In Figure \ref{fig-example}(b) the density estimation was calculated using Wand's algorithm. The bandwidth $\boldsymbol{H}$ was exactly the same as in Figure~\ref{fig-example}(a). It is easy to notice that the result is obviously inaccurate, as the results in Figures \ref{fig-example}(a) and \ref{fig-example}(b) should be the same. The density estimation depicted in Figure \ref{fig-example}(c) is for the diagonal bandwidth $\boldsymbol{H}$ (calculated using the \texttt{Hpi.diag\{ks\}} \textsf{R} function). The KDE is exactly the same, regardless of using Wand's algorithm or direct calculations. It is important to see that Figures \ref{fig-example}(b) and \ref{fig-example}(c) are very similar. This similarity suggests that  Wand's algorithm lose in some way most (or even all) the information carried by off-diagonal entries of the bandwidth matrix $\boldsymbol{H}$. In other words Wand's algorithm (in it's current form) is adequate only for constrained bandwidths.

\section{Problem identification and its solution}\label{sec-ident}

\subsection{The current form of the algorithm}

To compute the convolution (\ref{eq-bkde-conv}) (or optionally (\ref{eq-bkde-conv-L})) of two vectors the \emph{discrete convolution theorem} is used. However, this theorem requires two main assumptions about the two input vectors, that is $\boldsymbol{c}$ and $\boldsymbol{k}$. These wectors in signal processing's terminology are caled \emph{input signal} and \emph{impulse response}, respectively. The first assumption states that the two vectors must have the same length and the second assumption requires that the input signal be treated as a periodic one. The consequence of the above is that the so called \emph{zero-padding} and  \emph{wrap-around ordering} procedures are required. Details can be found for example in \citet[Chapter 13]{Press:1992}. In \cite{Wand-1994} the author suggests reshaping $\boldsymbol{c}$ and $\boldsymbol{k}$ as in (\ref{eq-K-2D}) and (\ref{eq-C-2D}). Here, for simplicity, only the two-dimensional variant is presented, as extensions to higher dimensions are straightforward. We have
\begin{align}\label{eq-K-2D}
\boldsymbol{k} =
\begin{bmatrix}
k_{0,0}   & k_{0,1}        & \cdots & k_{0,M_2}   &                & k_{0,M_2}   & \cdots        & k_{0,1} \\
k_{1,0}   & k_{1,1}        & \cdots & k_{1,M_2}   &                & k_{1,M_2}   & \cdots        & k_{1,1} \\
\vdots    & \vdots         & \ddots & \vdots      & \boldsymbol{0} & \vdots      & \ddots        & \vdots \\
k_{M_1,0} & k_{M_1,1}      & \cdots & k_{M_1,M_2} & \cdots         & k_{M_1,M_2} & \cdots        & k_{M_1,1} \\
          & \boldsymbol{0} &        & \vdots      & \boldsymbol{0} & \vdots      & \boldsymbol{0}& \\
k_{M_1,0} & k_{M_1,1}      & \cdots & k_{M_1,M_2} & \cdots         & k_{M_1,M_2} & \cdots        & k_{M_1,1} \\
\vdots    & \vdots         & \ddots & \vdots      & \boldsymbol{0} & \vdots      & \ddots        & \vdots \\
k_{1,0}   & k_{1,1}        & \cdots & k_{1,M_2}   &                & k_{1,M_2}   & \cdots & k_{1,1} \\
\end{bmatrix}
\end{align}
and
\begin{align}\label{eq-C-2D}
\boldsymbol{c} =
\begin{bmatrix}
c_{1,1}   & c_{1,2}        & \cdots & c_{1,M_2}   &       &                & \\
\vdots    & \vdots         & \ddots & \vdots      &       & \boldsymbol{0} & \\
c_{M_1,1} & c_{M_1,2}      & \cdots & c_{M_1,M_2} &       & \cdots \\
          & \boldsymbol{0} &        & \vdots      &       & \boldsymbol{0} & \\
\end{bmatrix}.
\end{align}
If one prefers to make use of the effective support property, as in (\ref{eq-bkde-conv-L}), $M_1$ and $M_2$ must be replaced by $L_1$ and $L_2$, respectively. However, this is only a technical procedure which does not affect the problem under consideration.

The sizes of the zero matrices are chosen so that after the reshaping of $\boldsymbol{c}$ and $\boldsymbol{k}$, they both have the same dimension $P_1 \times P_2, \times, \ldots, \times P_d$ (highly composite integers; typically, a power of 2). Now, to get the searched density estimate $\tilde{f}$ from (\ref{eq-bkde-conv}) we can apply the discrete convolution theorem, that is, we must do the following operations: 
\begin{align}\label{eq-disc-conv-theo}
\boldsymbol{C}&=F(\boldsymbol{c}), \;\;\; \boldsymbol{K}=F(\boldsymbol{k}), \;\;\;
\boldsymbol{S}=\boldsymbol{C}\boldsymbol{K}, \;\;\; \boldsymbol{s}=F^{-1}(\boldsymbol{S}) \end{align}
where $F$ stands for the Fourier transform and  $F^{-1}$ is its inverse. The sought density estimate $\tilde{f}$ corresponds to a subset of $\boldsymbol{s}$ in Eqn. (\ref{eq-disc-conv-theo}) divided by the product of  $P_1,P_2,\ldots,P_d$ (the so-called normalization), that is
\begin{align}
\tilde{f} = 
\frac{1}{P_1 \; P_2 \ldots P_d}
\boldsymbol{s} [1 : M_1, \ldots ,1 : M_d] 
\end{align}
where, for the two-dimensional case, $\boldsymbol{s}[a:b, c:d]$ means a subset of rows from $a$ to $b$ and a subset of columns from $c$ to $d$ of the matrix $\boldsymbol{s}$. 

Now we will try to discover what is wrong with $\boldsymbol{k}$ and $\boldsymbol{c}$ matrices, causing the problems described in Section \ref{sec-demo}. Wand's algorithm presented in \citet[appendix D]{Wand-1995} concerns only 1D case which works absolutely correct. In \cite{Wand-1994} the algorithm is generalized for higher dimensions. 
However, this generalization supports only constrained (diagonal) bardwidth matrices $\boldsymbol{H}$.
If we carefully look at (\ref{eq-K-2D}) it is easily to recognize that the wrap-around ordering used will support only kernels in orientations according to the coordinate axes, that is those where $\boldsymbol{H}$ is diagonal. If  $\boldsymbol{H}$ is unconstrained many entries in $\boldsymbol{k_l}$ of (\ref{eq-bkde-conv}) required to compute $\tilde{f}_{\boldsymbol{j}}$ does not occur in (\ref{eq-K-2D}). In other words entries for `negative times' (for example $k_{-1,2}$ or $k_{-1,-1}$) will not be recovered by the wrap-around ordering as  $k_{-1,2} \neq k_{1,2}$ and $k_{-1,-1} \neq k_{1,1}$ and so on. The above pairs of entries would be equal only if  $\boldsymbol{H}$ were diagonal. This implicitly explains why Figure \ref{fig-example}(b) is so similar to Figure \ref{fig-example}(c).

\subsection{The corrected algorithm}

Regarding the problems described in the previous subsection, we propose a different reshaping for $\boldsymbol{k}$ and $\boldsymbol{c}$ (now renamed to $\boldsymbol{k}_{new}$ and $\boldsymbol{c}_{new}$) which removes the problem presented in the paper. Note that now wrap-around ordering is \emph{not} utilized, only zero-padding is used. So, we have
\begin{align}\label{eq-K-2D-new}
\boldsymbol{k}_{new} =
\begin{bmatrix}
k_{-M_1,-M_2} & \cdots & k_{-M_1,0} & \cdots & k_{-M_1,M_2} & \\
\vdots        & \ddots & \vdots     & \ddots & \vdots       & \\ 
k_{0,-M_2}    & \cdots & k_{0,0}    & \cdots & k_{0,M_2}    & \boldsymbol{0} \\
\vdots        & \ddots & \vdots     & \ddots & \vdots       & \\ 
k_{M_1,-M_2}  & \cdots & k_{M_1,0}  & \cdots & k_{M_1,M_2}  & \cdots \\
              &        & \boldsymbol{0} &        & \vdots       & \boldsymbol{0} \\
\end{bmatrix}
\end{align}
and
\begin{align}\label{eq-C-2D-new}
\boldsymbol{c}_{new} =
\begin{bmatrix}
\boldsymbol{0} & \vdots     & \boldsymbol{0} & \vdots      & \boldsymbol{0} \\ 
\cdots     & c_{1,1}    & \cdots     & c_{1,M_2}   & \cdots \\
\boldsymbol{0} & \vdots     & \ddots     & \vdots      & \boldsymbol{0} \\
\cdots     & c_{M_1,1}  & \cdots     & c_{M_1,M_2} & \cdots \\
\boldsymbol{0} & \vdots     & \boldsymbol{0} & \vdots      & \boldsymbol{0} \\            
\end{bmatrix}
\end{align}
where the entry $c_{1,1}$ in Eqn.~(\ref{eq-C-2D-new}) is placed in row $M_1$ and column $M_2$. 
The sought density estimate $\tilde{f}$ corresponds to a subset of $\boldsymbol{s}$ in Eqn. (\ref{eq-disc-conv-theo}) divided by the product of  $P_1,P_2,\ldots,P_d$ (the so-called normalization), that is
\begin{align}
\tilde{f} = 
\frac{1}{P_1 \; P_2 \ldots P_d}
\boldsymbol{s} [(2M_1-1) : (3M_1-2), \ldots ,(2M_d-1) : (3M_d-2)]. 
\end{align}

As was mentioned in Section \ref{sec-algo}, $M_k$ values can be shrunk to some smaller values $L_k$. We propose to calculate ${L_k}$ using the following formula ($k=1, \cdots, d$)
\begin{align}\label{eq-L}
L_k
= 
\min 
\left( 
M_k-1, 
\text{ceiling}
\left( 
\frac{\tau \; \sqrt{ | \lambda | }}{\delta_k} 
\right)
\right)
\end{align} 
where $\lambda$ is the largest eigenvalue of $\boldsymbol{H}$ and $\delta_k$ is the mesh size from Eqn.~(\ref{eq-delta-k}). After some empirical tests we have found that $\tau$ can be set to around $3.7$ for a standard two-dimensional normal kernel.

After implementing the improved version of  Wand's algorithm (based on (\ref{eq-K-2D-new}) and (\ref{eq-C-2D-new})) and calculating density estimation analogous to that depicted in Figure \ref{fig-example}(b) we can easily conclude that now the plot is \emph{identical} to the one from Figure \ref{fig-example}(a). This implies that the weakness of the original algorithm being the main paper's subject was \emph{resolved}. Two dedicated \textsf{R} functions (\texttt{bkde.2D.no.fft.radial} and \texttt{bkde.2D.fft.radial.corrected}) are included as supplemental materials for the purpose of replication of Figure \ref{fig-example}. The latter implements the corrected Wand's algorithm.

\section{Conclusion}\label{sec-conlusion}
In the paper we have described a very serious problem of using FFT for calculation of multivariate kernel estimators when unconstrained bandwidth matrices are used. Next, we have discovered a satisfactory solution which rectifies the problem. As a consequence, the results given by direct evaluation of (\ref{eq-bkde-conv}) or by (\ref{eq-bkde-conv-L}) and by the proposed FFT-based algorithm based on (\ref{eq-K-2D-new}) and (\ref{eq-C-2D-new}) are identical for any form of the $\boldsymbol{H}$ bandwidth matrix. Our results can be used not only for direct KDE calculations, but also for calculation of a class of functionals which are very important for example in optimal bandwidth selection for KDE. Our results have been already implemented in the \pkg{ks} \textsf{R} package, starting from version 1.10.0.

\bibliographystyle{Chicago}

\bibliography{JCGS_SHORT_Gramacki}

\begin{thebibliography}{}

\bibitem[\protect\citeauthoryear{Andrzejewski, Gramacki, and
  Gramacki}{Andrzejewski et~al.}{2013}]{Andrzejewski:2013}
Andrzejewski, W., A.~Gramacki, and J.~Gramacki (2013).
\newblock Graphics processing units in acceleration of bandwidth selection for
  kernel density estimation.
\newblock {\em International Journal of Applied Mathematics and Computer
  Science\/}~{\em 23\/}(4), 869--885.

\bibitem[\protect\citeauthoryear{Duong}{Duong}{2015}]{ks}
Duong, T. (2015).
\newblock {\em Kernel Smoothing}.
\newblock \textbf{R}~package version~1.10.0.

\bibitem[\protect\citeauthoryear{Gramacki and Gramacki}{Gramacki and
  Gramacki}{2015}]{Gramacki:2015}
Gramacki, A. and J.~Gramacki (2015).
\newblock {FFT}-based fast bandwidth selector for multivariate kernel density
  estimation.
\newblock {\em arXiv.org preprint\/}.
\newblock \url{hhttp://arxiv.org/abs/1511.07482}.

\bibitem[\protect\citeauthoryear{Kulczycki and Charytanowicz}{Kulczycki and
  Charytanowicz}{2010}]{Kulczycki:2010}
Kulczycki, S. and M.~Charytanowicz (2010).
\newblock A complete gradient clustering algorithm formed with kernel
  estimators.
\newblock {\em International Journal of Applied Mathematics and Computer
  Science\/}~{\em 20\/}(1), 123--134.

\bibitem[\protect\citeauthoryear{Press, Flannery, Teukolsky, and
  Vetterling}{Press et~al.}{1992}]{Press:1992}
Press, W., B.~Flannery, S.~Teukolsky, and W.~Vetterling (1992).
\newblock {\em Numerical Recipes in C: The Art of Scientific Computing, Second
  Edition}.
\newblock Cambridge University Press.

\bibitem[\protect\citeauthoryear{Raykar, Duraiswami, and Zhao}{Raykar
  et~al.}{2010}]{Raykar:2010}
Raykar, V., R.~Duraiswami, and L.~Zhao (2010).
\newblock Fast computation of kernel estimators.
\newblock {\em Journal of Computational and Graphical Statistics\/}~{\em
  19\/}(1), 205--220.

\bibitem[\protect\citeauthoryear{Schauer, Duong, Gomes-Santos, and
  Goud}{Schauer et~al.}{2013}]{Duong:2013}
Schauer, K., T.~Duong, C.~Gomes-Santos, and B.~Goud (2013).
\newblock Studying intracellular trafficking pathways with probabilistic
  density maps.
\newblock {\em Methods in Cell Biology\/}~{\em 118}, 325--343.

\bibitem[\protect\citeauthoryear{Scott}{Scott}{1992}]{Scott:1992}
Scott, D. (1992).
\newblock {\em Multivariate Density Estimation: Theory, Practice and
  Visualization}.
\newblock Wiley.

\bibitem[\protect\citeauthoryear{Silverman}{Silverman}{1982}]{Silverman:1982}
Silverman, B. (1982).
\newblock {Kernel density estimation using the fast Fourier transform.
  Algorithm AS 176}.
\newblock {\em Applied Statistics\/}~{\em 31}, 93--99.

\bibitem[\protect\citeauthoryear{Silverman}{Silverman}{1998}]{Silverman:1998}
Silverman, B. (1998).
\newblock {\em Density Estimation for Statistics and Data Analysis}.
\newblock London: Chapman \& Hall/CRC.

\bibitem[\protect\citeauthoryear{Simonoff}{Simonoff}{1996}]{Simonoff:1996}
Simonoff, J. (1996).
\newblock {\em Smoothing Methods in Statistics}.
\newblock Springer Series in Statistics.

\bibitem[\protect\citeauthoryear{Wand}{Wand}{1994}]{Wand-1994}
Wand, M. (1994).
\newblock Fast computation of multivariate kernel estimators.
\newblock {\em Journal of Computational and Graphical Statistics\/}~{\em
  3\/}(4), 433--445.

\bibitem[\protect\citeauthoryear{Wand and Jones}{Wand and
  Jones}{1995}]{Wand-1995}
Wand, M. and M.~Jones (1995).
\newblock {\em Kernel Smoothing}.
\newblock Chapman \& Hall.

\bibitem[\protect\citeauthoryear{Wand and Ripley}{Wand and
  Ripley}{2015}]{KernSmooth}
Wand, M. and B.~Ripley (2015).
\newblock {\em Functions for Kernel Smoothing Supporting Wand \& Jones (1995)}.
\newblock \textbf{R}~package version~2.23-15.

\end{thebibliography}
\end{document}